# Reversible Computing with Fast, Fully Static, Fully Adiabatic CMOS

Michael P. Frank
Cognitive & Emerging Computing Department
Sandia National Laboratories
Albuquerque, NM, USA
mpfrank@sandia.gov

Robert W. Brocato
RF Microsystems Department
Sandia National Laboratories
Albuquerque, NM, USA
orcid:0000-0001-9751-1234

Brian D. Tierney
Radiation Hard CMOS Technology Dept.
Sandia National Laboratories
Albuquerque, NM, USA
bdtiern@sandia.gov

Nancy A. Missert
Nanoscale Sciences Department
Sandia National Laboratories
Albuquerque, NM, USA
orcid:0000-0003-2082-2282

Alexander H. Hsia
Sandia National Laboratories
Albuquerque, NM, USA

*Requiescat in pace*

*Abstract*— To advance the energy efficiency of general digital computing far beyond the thermodynamic limits that apply to conventional digital circuits will require utilizing the principles of reversible computing. It has been known since the early 1990s that reversible computing based on adiabatic switching is possible in CMOS, although almost all of the "adiabatic" CMOS logic families in the literature are actually not fully adiabatic, which limits their achievable energy savings. The first CMOS logic style that achieved truly, fully adiabatic operation if leakage was negligible (CRL) is not fully static, which leads to a number of practical engineering difficulties in the presence of certain nonidealities. Later, "static" adiabatic logic families were described, but they were not actually fully adiabatic, or fully static, and were much slower.

In this paper, we describe a new logic family, *Static 2-Level Adiabatic Logic* (S2LAL), which is, to our knowledge, the first CMOS logic family that is both *fully* static, and *truly, fully* adiabatic (modulo leakage). In addition, S2LAL is, we think, the *fastest possible* such family (among fully pipelined sequential circuits), having a latency per logic stage of one "tick" (transition time), and a minimum clock period (initiation interval) of 8 ticks. S2LAL requires 8 phases of a trapezoidal power-clock waveform (plus constant power and ground references) to be supplied. We argue that, if implemented in a suitable fabrication process designed to aggressively minimize leakage, S2LAL should be capable of demonstrating a greater level of energy efficiency than *any* other semiconductor-based digital logic family known today.

## I. INTRODUCTION

Continuing to improve the computational energy efficiency of conventional CMOS technology is becoming increasingly difficult, due to numerous physical effects that emerge as device dimensions shrink and per-area densities increase [1]. This is, however, unsurprising, since the laws of physics guarantee that the energy efficiency of the conventional *non-reversible* computing paradigm is fundamentally limited [2][3]. But, an alternate paradigm for general digital computing called *reversible computing* [4][5] offers an *alternative* path for improving computer energy efficiency which bypasses the many barriers to further improvement that the conventional approach faces.

Although the reversible approach imposes a number of overheads, we can foresee that, in the long run, as the underlying technology improves, ultimately it must win out, as the benefits will eventually come to outweigh the overheads [6][7]. Maximizing the benefits will likely require new classes of devices that exploit quantum phenomena to minimize dissipation as a function of delay [8], but we can begin to gain experience today with the issues and principles of reversible design by prototyping reversible circuits in CMOS. As manufacturing processes move into the 3$^{rd}$ dimension and cost per device continues to decrease, even implementations of reversible computing that are based on CMOS could come to outperform conventional approaches, in terms of aggregate parallel performance within such system constraints as size, weight, cost, and power.

The fact that reversible computing can be implemented using a fully adiabatic style of sequential CMOS was first shown by Younis and Knight in 1993, in their Charge Recovery Logic (CRL) family [9]. However, CRL is a *dynamic* logic family, *i.e.*, some circuit nodes are left floating in some circumstances. This leads to various design difficulties which we will discuss. To avoid those problems, there is a need for a *fully static* family of adiabatic CMOS. Younis described a static version of his SCRL (split-level CRL) family in 1994 [10], but it was relatively slow, requiring 24 adiabatic transition times ("*ticks*") per clock cycle. Also, the original version of SCRL was not quite fully adiabatic either [7], or even fully static—it had some floating nodes.

We would thus like to have an adiabatic CMOS logic family that is both fully static *and* fully adiabatic, and further that optimizes the following key performance characteristics:

*1) Minimal latency:* The number of ticks of latency per layer of logic should be as small as possible (1 tick).

This work is supported by the Advanced Simulation and Computing (ASC) program at the U.S. Department of Energy's National Nuclear Security Administration (NNSA). Sandia National Laboratories is a multi-mission laboratory managed and operated by National Technology and Engineering Solutions of Sandia, LLC, a wholly owned subsidiary of Honeywell International, Inc., for NNSA under contract DE-NA0003525. This document describes objective technical results and analysis. Any subjective views or opinions that might be expressed in this document do not necessarily represent the views of the U.S. Department of Energy or the United States Government. Approved for public release, SAND2020-9140 O.

*2) Maximum throughput:* The number of ticks per clock period (initiation interval) should also be minimal.

In this paper, we describe a candidate for this, a fully static and fully adiabatic CMOS logic family that we call S2LAL, for "static, 2-level adiabatic logic," that requires only 8 ticks per clock period (*i.e.*, 3× faster than static SCRL), which we think is minimal for a fully static, fully adiabatic CMOS logic.

The remainder of this paper is organized as follows: In Sec. II, we describe in more detail our motivation for developing a fully static adiabatic CMOS logic family, and briefly describe the limitations of the existing approaches, and the requirements that a correct solution needs to meet. In Sec. III, we describe S2LAL, and explain its operation in some detail. In Sec. IV, we summarize ideas for future work and conclude.

## II. REQUIREMENTS FOR FULLY STATIC ADIABATIC CMOS

In this section, we review our motivation for designing an adiabatic CMOS logic family that is also fully static, explain the reasons why existing adiabatic CMOS logic families are not yet fully satisfactory, and we set forth formal technical requirements that any fully satisfactory solution will need to meet.

*A. What is fully static operation, and why is it needed?*

Briefly, when we call a particular design discipline *fully static*, this means that every node in the circuit is at all times connected to some external voltage reference, either directly via a low-impedance path, or at least indirectly, through turned-on switches of moderate impedance (*e.g.*, on the order of 1–10 kΩ is typical for a turned-on logic FET). Implicitly, there should also not exist competing paths of comparable impedance to sources at *different* reference levels, since this would imply a substantial "short circuit" current with associated power dissipation.

This concept of "static" is the gist of what is meant when we talk about "static CMOS" or "static RAM" (SRAM), in contrast to *dynamic* design families for logic and memory, where circuit nodes are allowed to *float* at a dynamic level (*i.e.*, without low-impedance paths to any sources). The voltage on a floating node can spontaneously vary over time, in various ways, including:

*1) Voltage drift*, due to leakage (*e.g.*, subthreshold leakage or gate leakage) to/from adjacent nodes in the circuit;

*2) Voltage "sag,"* due to capacitive voltage-divider effects from parasitic capacitances with nearby nodes in the circuit;

*3) AC coupling* via capacitive paths to nearby nodes, which become low-impedance at high frequencies.

Whatever their cause, deviations in the dynamic voltage level of a floating node can lead to various difficulties in an adiabatic circuit, including the following:

*1) Sudden energy dissipation* (which we informally term "sparking") when the node is reconnected to a reference source; if the node's capacitance is $C$, and its voltage disparity from the reference level is $\Delta V$, then the energy dissipated is $C(\Delta V)^2/2$. Such sparking events are particular anathema for applications needing *low and steady* instantaneous power for privacy, *e.g.*, in the context of an adversary analyzing EM emissions.

*2) Logical errors* in the case of circuit nodes whose voltage deviation becomes large enough to affect the digital behavior of the circuit. This can easily occur in any scenario where the clock frequency is slowed or paused for any reason, or even at relatively high frequencies, if a node remains floating for sufficiently many cycles due to low activity in that part of the circuit; *e.g.*, a DRAM cell will go invalid if not refreshed.

In contrast, in a fully static, fully adiabatic CMOS logic, power dissipation will be low and relatively steady, and logical errors will be essentially impossible, no matter how low the clock speed, even if the clocks are paused completely for system-level purposes (*e.g.*, flow control, standby mode, or diagnostics).

Thus, we can see that there are significant advantages to be gained, in terms of both efficiency and usability, from the development of fully static versions of truly, fully adiabatic logic.

Due to these many advantages, we will henceforth use the phrase "*perfectly adiabatic*" as a technical term referring to a CMOS circuit that is simultaneously fully static, and truly, fully, adiabatic. In such circuits, minimum energy dissipation is only limited by leakage, and can be made as low as desired if the underlying technology is refined to reduce leakage sufficiently.

*B. Limitations of some existing adiabatic CMOS families.*

To our knowledge, no previously published description of an adiabatic CMOS logic family for general sequential digital computing quite achieves our goal of "perfectly adiabatic" operation. Here we briefly review how some of the previously described adiabatic or quasi-adiabatic logic styles depart from this ideal.

Early efforts in the direction of adiabatic CMOS from 1978–1992 by Fredkin & Toffoli [11], Seitz *et al.* [12], Koller & Athas [13], Hall [14], and Merkle [15] were incomplete, either not qualifying as fully adiabatic, or not illustrating general sequential operation. As mentioned earlier, Younis and Knight's CRL [9], the first fully adiabatic sequential CMOS logic family, was not static. Athas *et al.* attempted to describe an 8 tick/cycle static adiabatic logic family in 1994 [16], but that design was in fact neither fully adiabatic nor fully static (having some internal nodes that are left floating for part of the cycle, and non-adiabatically overwritten). Younis [10] described a 24-tick static variant of SCRL, but it was not quite fully adiabatic, due to some transistors that gradually turn themselves off as a threshold level is approached. And, neither was it fully static, since it, too, includes internal nodes that become isolated in some cases. However, both of these problems with static SCRL would be eliminated if a modification described in [7] is applied.

In any case, the novel perfectly adiabatic CMOS logic family described in this document (S2LAL) offers the following advantages versus even a *repaired* version of static SCRL:

*1) 8× lower latency*, 1 tick of latency per layer of logic in S2LAL, as opposed to 8 ticks/layer in static ("3 phase") SCRL;

*2) 3× higher throughput*, due to a 3× shorter clock period, specifically 8 ticks/cycle in S2LAL, as opposed to 24 ticks/cycle in static SCRL;

*3) 3.75× fewer distinct clock waveforms;* that is, only 8 distinct clocks, as opposed to 30 in (single-rail) static SCRL.

S2LAL has one disadvantage versus static SCRL: Namely, it is a quad-rail logic family, and thus requires more transistors in total per logic stage than static SCRL, which can operate as a single-rail logic. But this advantage of SCRL is at least somewhat negated by the fact of its longer clock period, which inflates both the minimum energy dissipation per device cycle, and the hardware cost to attain any required aggregate level of throughput. Finally, the large number of clock rails required by static SCRL can be expected to contribute substantially to its layout area, and its long clock period makes the design of high-quality energy-recovering resonant oscillators, vital for achieving good system-level energy efficiency, more difficult. However, a much more detailed design study would be needed to directly compare S2LAL against static SCRL for use in specific applications in specific fabrication processes. For now, we just focus on giving a careful description of the new logic style.

*C. Requirements for Fully Static, Fully Adiabatic Operation*

The requirements for truly, fully adiabatic operation (when ignoring off-state leakage) have been described elsewhere [16] and we summarize them only briefly here:

*1) No diodes.* Diodes must be avoided in the main paths for logic node charging and discharging, due to their built-in voltage drop.

*2) No "spark" events.* This means, do not turn a switch (*e.g.*, a FET) "on" when there is a non-negligible voltage difference between its source and drain terminals. More generally, do not rapidly produce a large voltage change by any means.

*3) No "squelch" events.* This means, do not turn a switch "off" when there is a non-negligible current between its source and drain terminals (unless there is an alternative, comparably low-impedance path for that current).

*4) Quasi-trapezoidal power-clocks.* Supply voltage waveforms should rise and fall gradually, and should have flat tops and bottoms. This rule is implied by the previous two.

To guarantee fully static operation in the presence of these constraints, we must ensure that every internal circuit node (*i.e.*, every circuit node that is not directly connected by a low-impedance path to an externally supplied voltage reference) must *at all times* be, at least, *less*-directly connected to a voltage reference (that is, via a medium-impedance path through turned-on devices) in one of the following manners:

*1) Connected via a medium-impedance path through turned-on transistor(s) to a single constant-voltage reference;*

*2) Connected via a medium-impedance path through turned-on transistor(s) to a single variable-voltage reference;*

*3) Connected in a way that is actively transitioning (in either direction) between conditions 1 & 2 above, with one path in the process of being connected while the other is in the process of being disconnected, and with at least one path having no more than medium impedance at all times throughout the transition, and where, throughout the transition period, the level of the variable-voltage reference in question is being held constant at the same level as the constant-voltage rail;*

*4) Connected in a way that is similarly actively transitioning between two different paths to a single supply reference (whether it is constant-voltage or variable-voltage).*

Obeying the above rule ensures that no appreciable voltage sag or drift can ever occur on any node in the circuit, since, at all times, every node enjoys at least one path of no more than medium impedance to a reference node at its nominal voltage level.

To be slightly more specific, although the following values are somewhat arbitrary, for purposes of this document we will consider "medium impedance" to refer to charging paths having ~100–100,000 Ω impedance in the DC limit. We will assume that, in any "ordinary" CMOS process, a minimum-width MOSFET will typically have an impedance value within this range whenever it is considered to be in the "on" state (*e.g.*, when $V_{GS} > V_{th}$ in the case of an nFET), and will ideally have some much larger impedance value (but at minimum, >1 MΩ) whenever it is in the "off" state (*e.g.*, $V_{GS} = 0$).

This concludes our discussion of the requirements that we consider necessary for achieving what we call "perfectly adiabatic" operation. With these requirements in mind, careful inspection of a given logic family can allow the reader to verify whether it qualifies. We now proceed to describe S2LAL, which we conjecture is in fact the fastest possible qualifying family.

III. DESCRIPTION OF THE S2LAL LOGIC FAMILY

In this section, we give a detailed technical description of the new S2LAL logic family. We start by presenting a concise formal notation for clock and data signals, then describe the basic circuit elements of S2LAL in a hierarchical fashion.

*A. Important Notations*

A semi-formal mathematical notation which will be helpful for describing signals (both clock and data signals) will now be introduced. As the name of S2LAL suggests, we will use just two nominal voltage levels, 0 (low) and $V_{dd} \gtrsim 2|V_t|$ (high) (where $|V_t|$ is the absolute FET threshold), in contrast to SCRL, which also uses a third level $V_{dd}/2$. As previously suggested, we divide time into discrete intervals we call *ticks*, each of the same duration $\overline{\tau}_{tr}$, which is also the *maximum* time allowed for any adiabatic transition between nominal levels, driven by the rise (or fall) of an externally supplied voltage reference between the levels over some actual period $\tau_{tr} \leq \overline{\tau}_{tr}$. These transitions ideally have a linear ramp shape; if not exactly linear, dissipation per transition is multiplied by a shape factor $\xi_{tr}$. To first order (ignoring resistance and capacitance variations, wire inductances, and exponential tails at the start and end of the ramp), the active dissipation from adiabatically driving a load $C_L$ between nominal levels (in either direction) is then

$$E_a = \xi_{tr} C_L V_{dd}^2 \frac{RC_L}{\tau_{tr}}, \qquad (1)$$

where $R$ is the effective DC resistance of the charging path. This classic expression (*cf.* [16]) illustrates that, setting aside losses from leakage, the energy dissipation due to active transitions can be made as low as desired, as the ramp time $\tau_{tr}$ is made longer.

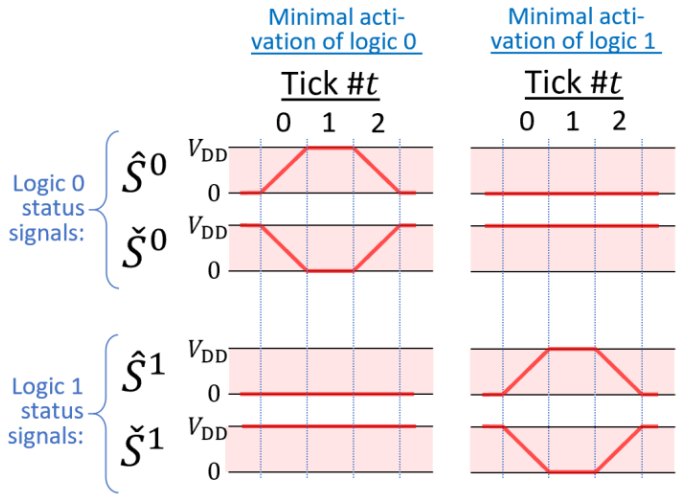

Fig. 1. Example of signal waveforms on a quad-rail bus for representing the abstract logic signal $S$. In this example, the bus stays valid for only one tick. This abstract signal could be denoted $S_{0,2}$ since it goes valid on tick #0 and invalid on tick #2. Note the active-high signals $\hat{S}^L$ pulse high when active, and the active-low signals $\check{S}$ pulse low. There is a separate signal pair to convey each possible logic symbol, and no two symbols are simultaneously active.

Within a given operation sequence for the circuit, we identify consecutive ticks using respective integers $t = 0, 1, 2, ...$, for example. For periodic AC supplies, the clock period is taken to be an integral number of ticks, $\tau_p = n\bar{\tau}_{tr}$. Thus, the clock frequency is $f = 1/\tau_p = (n\bar{\tau}_{tr})^{-1}$.

In S2LAL (as with other quad-rail styles, including CRL and 2LAL [18]), different possible logic values (symbols) are represented by independent signal pairs. Ordinarily, we use binary logic with two possible symbols $L = 0,1$. However, multi-valued logics are easily supported as well (*e.g.*, ternary logic would use six signals). In general, to represent any of $k$ distinct logic symbols, we can use a bus of $2k$ signals.

For any given logic symbol $L$ (*e.g.*, logic 1), during any given tick when the two signals corresponding to that symbol are being held at nominal levels, those signals can be in either of two possible conditions: Namely, they are either, together, *actively representing $L$* ("*active*"), or *not actively representing $L$* ("*inactive*"). During other ticks, the two signals simultaneously transition (in either direction) between active and inactive states.

Further, the two signals making up a given signal pair corresponding to a logic symbol $L$ must always have *complementary* levels: That is, one is high whenever the other is low, and vice-versa. We say that one of the two signals is *active-high* (*i.e.*, it is high when $L$ is being actively represented), and the other one is *active-low* (*i.e.*, low when $L$ is being actively represented). Normally, at most one of the $n$ various alternative symbols that can be represented on a given signal bus may be active at any given time.

We will use the following notation, and simplified variations thereof, to denote the signal pair for any given logic symbol $L$.

$$S^L_{t_b,t_e} = (\hat{S}^L_{t_b,t_e}, \check{S}^L_{t_b,t_e}) \quad (2)$$

Here, $S$ can be any letter representing the name and/or semantic purpose of this overall signal bus within the larger architecture. The "hat" (^) and "cup" (ˇ) accents mark the active-high and active-low signals in the pair, respectively. The superscript $L$, if present, indicates the logic symbol being represented; if absent, it is implicit and/or unimportant. The subscripts $t_b$ and $t_e$, if present, are the "begin" and "end" ticks of the *valid* period for the symbol; that is, the signal may transition from inactive to active in tick $t_b$, and (if it did) then it will transition back to inactive in tick $t_e$. Usually, the active period is given in context and so the second subscript $t_e$ may be omitted.

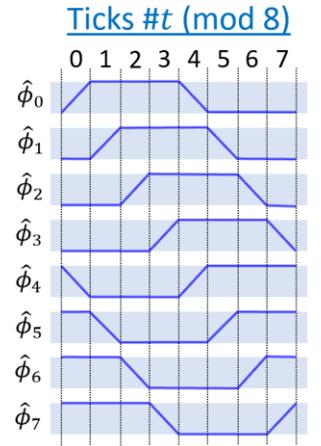

Fig. 2. The active-high clock signals $\hat{\phi}_i$ for the 8 phases of S2LAL. (The 8 corresponding active-low signals are relabelings of these.)

The presence of the symbol $L$ in some period is represented by a temporary pulse on its signal pair, in which $\hat{S}^L$ pulses high and $\check{S}^L$ pulses low. Fig. 1 illustrates, for the ordinary case with $k = 2$ logic symbols and a minimum valid period of 1 tick.

Our notation for periodic clock signals will be the same, but with $S$ being $\phi$, and stipulating that in the context of an $n$-tick clock period, the beginning and end tick numbers $t_b$ and $t_e$ shall always be considered to be taken modulo $n$. In other words, any values of $t_b$ and $t_e$ are implicitly mapped down to the range $0, 1, ..., n-1$, and they denote that the clock signal will transition to active during all ticks $t \equiv t_b \pmod{n}$, and to inactive in all ticks $t \equiv t_e \pmod{n}$.

Note a simple optimization applies given symmetrical duty cycles, *i.e.*, if $n$ is even and $t_e - t_b \equiv n/2 \pmod{n}$, by leveraging the observation that, in this case, $\hat{\phi}_i = \check{\phi}_{i+n/2}$; thus, $n$ distinct clock networks suffice to distribute both active-high and active-low clock signals for all $n$ phases $0 \leq i < n$.

In S2LAL, the clock period is $n = 8$ ticks, and $t_e - t_b \equiv 4 \pmod{8}$, so we can use this optimization, and require only 8 clock distribution networks to provide 8 pairs $\phi_i = (\hat{\phi}_i, \check{\phi}_i)$ of complementary clock signals covering all 8 of the possible timing phases. See Fig. 2 above for an illustration.

We now describe the basic circuit elements of S2LAL.

*B. CMOS Transmission Gates*

S2LAL, like the other 2-level adiabatic logic styles before it (such as 2LAL and CRL), makes extensive use of CMOS transmission gates. Because we use these so widely, we introduce a simplified graphical notation for them; see Fig. 3 below.

It's important to note that each FET device in the transmission gate *must* be a 4-terminal device, having an implicit body or back-gate terminal (not shown) which *shall not* be connected to either channel terminal of the FET (what would normally be called the "source" and "drain" terminals; but, in bidirectional T-gates, such labels should be avoided, since the source/drain roles will be alternating). Instead, the body (or back-gate) terminal of each nFET should be connected to GND (fixed 0 V supply), and that of each pFET to a fixed logic high reference, $V_{\text{dd}}$.

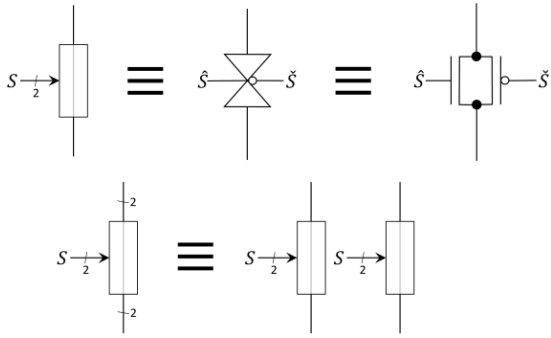

Fig. 3. (Top) Simplified icon for CMOS transmission gates (T-gates). The control input is implicitly a pair of complementary active-high and active-low signals. The active-high signal controls the gate of the nFET and the active-low signal controls the gate of the pFET. Thus, when the input pair is "active," the T-gate is turned "on" (conducting). (Bottom) When the T-gate icon is depicted with an $n$-rail parallel bus on its channel terminals, this just implies that we have $n$ parallel T-gates with identical controls.

Alternatively, one may also bias the body terminal levels by an additional offset ($+V_B$ for nFETs and $-V_B$ for pFETs), which will effectively increase the absolute device threshold, if desired, towards a larger magnitude (where the precise increase depends on device geometry); this will reduce off-state subthreshold currents. Note, however, that if we do this, in order to retain decent on-state conductance of transmission gates over the entire signal range $[0, V_{\text{dd}}]$, we must still respect the constraint $V_{\text{dd}} \gtrsim 2|V_t|$, where $V_t$ is now the *body-effected* threshold. Also, it's important to be cognizant that reducing subthreshold leakage may not reduce total off-state leakage by very much, and thus may not be worthwhile, in cases where the total leakage is already dominated by tunneling current through the gate oxide. Detailed techniques to reduce gate leakage are a topic for a later paper.

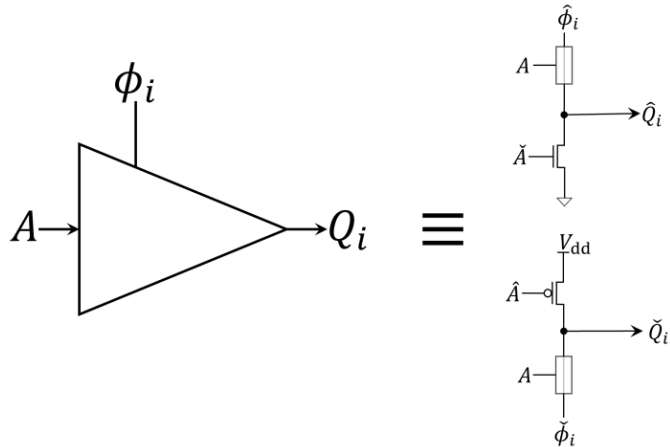

Fig. 4. Unlatched static adiabatic buffer. It takes a (dual-rail) input $A = (\hat{A}, \check{A})$ and is driven by an externally-supplied (dual-rail) clock pair $\phi = (\hat{\phi}, \check{\phi})$ which goes active on ticks #$i$ (mod 8) and inactive on ticks #$i + 4$ (mod 8). An implicit constraint on $A$ is that it *must be stable throughout the active pulse on $\phi$*, which means (given that the full pulse on $\phi$, including transitions, takes 5 ticks) that $A$ must go valid for at least 5 ticks, so a complete pulse on $A$ takes up 7 out of the 8 ticks in the cycle (Fig. 7). In the 8th tick, $A$ rests in the inactive state.

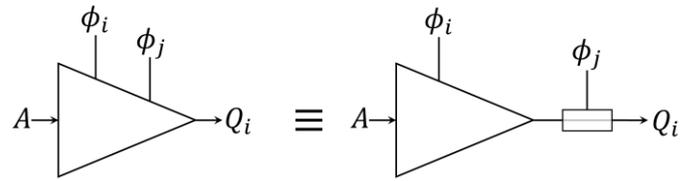

Fig. 5. *Latching* static adiabatic buffer, created by inserting a T-gate on each output line from the unlatched buffer. This structure requires two clocks, in phases $i$ and $j$, where $j$ is different from both $i$ and $i + 4$ (mod 8). If $j$ is any of $i - 1, i - 2, i - 3$, then the T-gate will be active (conducting) while $\phi_i$ is going active, but will be inactive (non-conducting) while $\phi_i$ is going inactive, thereby leaving $Q_i$ latched in a valid state. If $j$ is any of $i + 1, i + 2, i + 3$, the identical structure can be used to *decompute* a previously valid output; we will see how to use this in Sec. III.E.

*C. Unlatched Static Adiabatic Buffers*

Armed with FETs and transmission gates, as we described above, we can now construct the most basic element of S2LAL, which we call the (unlatched) *static adiabatic buffer* (see Fig. 4). This circuit was first described by Athas *et al.* in 1994 [16], although called an *adiabatic amplifier* in that paper.

The unlatched static adiabatic buffer takes just 6 transistors, and its mechanism of operation is easy to follow. Its dual-rail (complementary) output $Q = (\hat{Q}, \check{Q})$ is driven by a corresponding pair $\phi = (\hat{\phi}, \check{\phi})$ of clock rails, which hold at active for a period $t_{\phi\text{act}}$ (in our case, 3 ticks). Its dual-rail input signal $A = (\hat{A}, \check{A})$ is constrained by specification to remain stable (whether in an active, or inactive state) throughout this driving pulse, including its begin and end transitions, thus for at least 5 ticks. So $A$ can only transition during the other 3 ticks of the cycle, which is the period when $\phi$ is inactive. Note that whenever $A$ is in the inactive state, $Q$ is tied by single FETs to the rails holding *it* in the inactive state, whereas if $A$ is in the active state, $Q$ is tied to the clock pulse through the transmission gate. But, when $A$ transitions (goes active or inactive), we have $\phi = (0, V_{\text{dd}})$ at that time, and therefore we're only reconnecting $Q$ between two references at identical levels, which obeys the Sec. II.C rules.

We will show how to generalize this basic buffer structure to create more general adiabatic logic gates later, in Sec. III.F.

But first, an obvious question to ask about the buffer structure is: Given that the input has to remain valid for two ticks longer than the output, how can we chain such gates together while still maintaining a constant clock period? This is solved (similarly to static SCRL) using *latching logic* and careful sequencing of handoffs, described in the next two subsections.

*D. Latching Static Adiabatic Buffers*

The first step in fleshing out S2LAL as a full sequential logic style is to turn the basic static adiabatic buffer from the previous subsection into a *latching* buffer. To do this we simply run its output through another T-gate, controlled by an out-of-phase clock $\phi_j$ (Fig. 5). As long as $\phi_j$ is already active when $\phi_i$ goes active, this will have the effect of latching the output $Q$, so that it will remain at a valid level after $\phi_i$ goes inactive. To prevent this from leading to a *dynamic* (rather than fully static) logic family, while also maintaining fully adiabatic operation, requires further structure, which we will discuss in the next subsection.

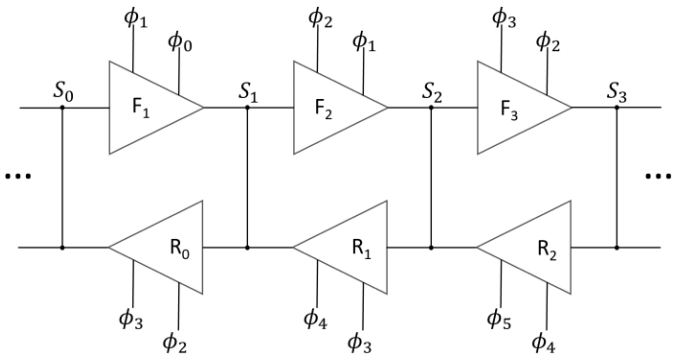

Fig. 6. Basic structure of a reversible pipeline in S2LAL. This one comprises a shift register, but pipelines of arbitrary logical complexity follow the same basic timing pattern. Each forward stage $F_i$ for computing an output signal $S_i$ from an input $S_{i-1}$ is paired with a corresponding reverse stage $R_{i-1}$ which *decomputes* $F_i$'s input signal $S_{i-1}$ from the output $S_i$. Note each $F_i$ is enabled (connected to its output) in phase $i-1$ and activated (computing its output) in phase $i$. Meanwhile, each reverse gate $R_i$ is activated (matching its output $S_i$) in phase $i+2$, and then enabled (connecting it to its output) in $i+3$, at the same time that $F_i$ is being disabled (its output terminal is going high-impedance). Thus, $S_i$ is, at every moment, connected to some source. Later, in $i+6$, as soon as $R_{i-1}$ is no longer actively using signal $S_i$, it gets decomputed by $R_i$, and then in $i+7$, control of $S_i$ switches back to $F_i$ in preparation to initiate another cycle on $i+8$. Careful inspection shows that there is no slack in this cycle; *i.e.*, it cannot be shortened at all without violating our requirements.

### E. Reversible Pipeline Sequencing

To permit arbitrary sequential logic while retaining fully adiabatic (and fully static) operation requires a *reversible pipeline*. An example of this (for a simple shift register) is shown in Fig. 6, with corresponding signal timing diagrams in Fig. 7.

As is usual for fully adiabatic circuits, such a pipeline consists of paired forward and reverse stages. In the general (multi-signal) case, if a forward stage $F_i$ computes some particular invertible function $\vec{S}_i = G_i(\vec{S}_{i-1})$ of its input vector $\vec{S}_{i-1}$, then the matching reverse stage $R_{i-1}$ must be designed to reconstruct (a copy of) the input of $F_i$ using $G_i^{-1}$, the inverse function to $G_i$. *I.e.*, $R_{i-1}$ reconstructs a copy of $\vec{S}_{i-1} = G_i^{-1}(\vec{S}_i)$, and then $R_{i-1}$ is used to *decompute* $\vec{S}_{i-1}$ *through* its copy.

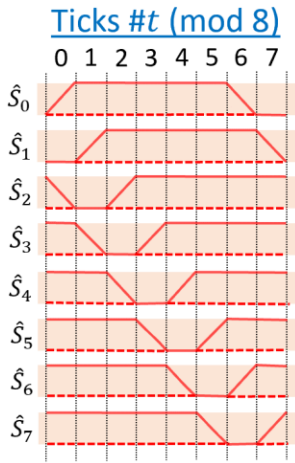

Fig. 7. The valid periods for signals $S_i$ computed on phases $i$ of an S2LAL pipeline. (Dotted lines show the level if the signal is not activated in the given cycle.)

Note that each signal stays valid (in a data-carrying state) for 5 ticks out of the 8 and is invalid (carrying no data) for 1 tick. During the middle tick of the valid period, and during the invalid tick, control of the signal is handed off between the forward and reverse gates that drive it.

The reason why each signal $S_i$ must stay valid for $\geq 5$ ticks (thus why the complete clock period must be $\geq 8$ ticks) is that before the handoff from $F_i$ to $R_i$ can occur, two transitions must occur, in sequence: Namely, the next forward stage $F_{i+1}$ must compute *its* output $S_{i+1}$, then $R_i$ must use that to reconstruct a copy of $S_i$ at its (tristated) output port before its output T-gate is enabled. Then, we need another 2-step sequence *after* the hand-off, while decomputing $S_{i-1}$ from $S_i$, before $S_i$ itself can finally be decomputed by $R_i$. This is why we posit that S2LAL is the *fastest possible* perfectly adiabatic (*i.e.*, fully static and fully adiabatic) pipeline technique for reversible CMOS.

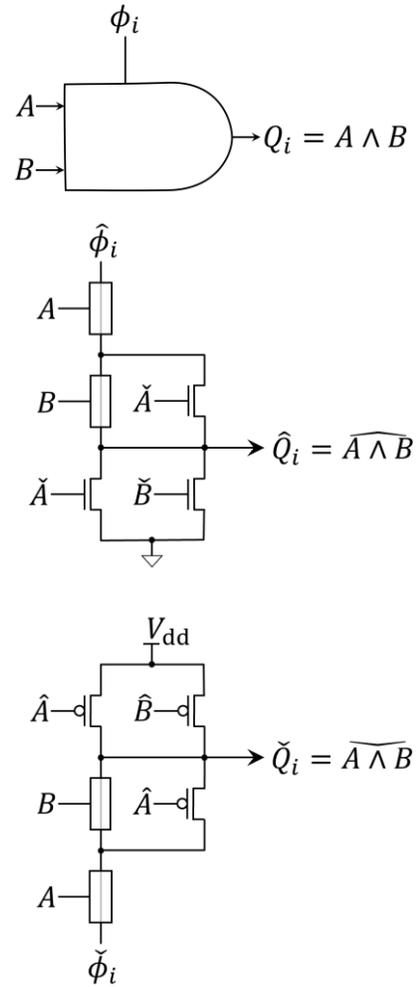

Fig. 8. S2LAL static AND gate icon (top) and circuit (bottom).

### F. Logic Functions in S2LAL

Implementing general logic in S2LAL is of course more involved, but still doable. The T-gates in Fig. 4 are replaced with more generalized *activation networks*, and the individual FETs in Fig. 4 are replaced with more general *hold-inactive networks*. Both networks must be carefully designed so as to never provide a path to both supplies simultaneously, except when the input is switching, since then the clock matches the constant supply. We also must take care to ensure that no internal nodes are ever left floating; there must always be a path from each internal node to at least one supply. These rules imply the minimal structure of each network, given its logical function. Minimal circuits for AND/OR functions are shown in Figs. 8 and 9.

The S2LAL unlatched AND gate takes 14 transistors. For a latching version, one simply inserts a T-gate in each output line, as with the buffer. Note that the internal nodes can never float, because at every moment there is a path from each internal node to either a constant or a variable supply reference (when input A is inactive, the internal node is held inactive as well).

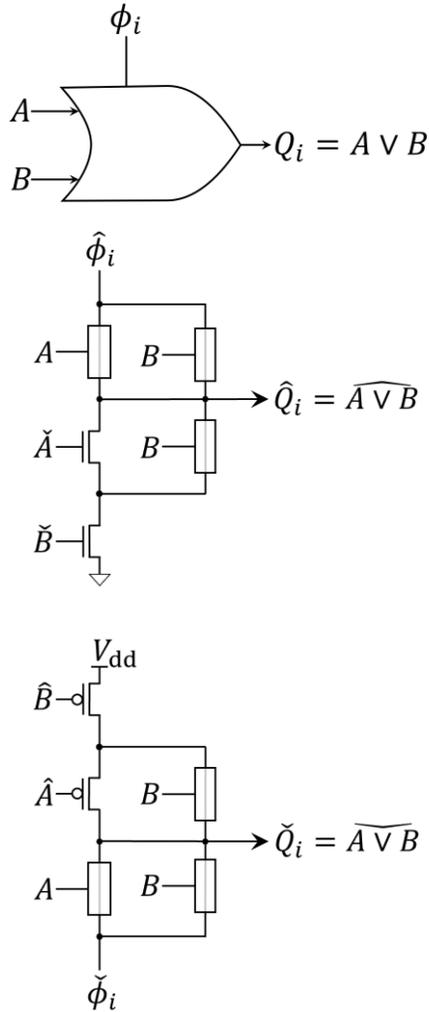

Fig. 9. S2LAL static OR gate icon (top) and circuit (bottom).

The unlatched OR gate takes 16 transistors. Note that two more transistors are required because when both inputs are active, we require a full transmission gate to tie the internal node of the hold-inactive network to the clock to ensure it does not float.

Inverting gates such as NOT, NAND and NOR are straightforward using quad-rail signaling, since pulsed complementary (active-high and -low) signal pairs for the *logical* complements of any given signal are also available. That is, if the signal pair we've been calling "$A$" is shorthand for $A^1$ (meaning $A = 1$), then there is also a signal pair $A^0$ (meaning $A = 0$) which pulses active on a given cycle if and only if $A^1$ does not, and similarly for $B$. So, we can implement NOT/NAND/NOR gates using the buffer and AND/OR gates, thusly (invoking DeMorgan):

$$\text{NOT}(A^1) = \text{BUFFER}(A^0) \tag{3}$$

$$\text{NAND}(A^1, B^1) = \text{OR}(A^0, B^0) \tag{4}$$

$$\text{NOR}(A^1, B^1) = \text{AND}(A^0, B^0) \tag{5}$$

Note if we wish to always maintain the quad-rail invariant that, for any signal pair $S^L$ representing a logic symbol $L = 0,1$, a logically complementary signal pair $S^{1-L}$ is also available, which pulses active on all and only those cycles that $S^L$ does not (as illustrated in Fig. 1), then we have to ensure that our design includes additional gates as needed to compute those signals. But if subsequent logic does not require the logical complements of all signals, this extra complexity may be reduced.

We should also note that the need to utilize invertible functions at each stage of the reversible pipeline in general adds an additional layer of complexity, but this gets into a higher-level topic, regarding the design and analysis of reversible algorithms (realized here as sequential pipelines). We consider that topic to be beyond the scope of this particular paper.

This concludes our presentation of S2LAL.

## IV. FUTURE WORK AND CONCLUSION

In this section, we briefly discuss some ideas for future work and conclude the paper.

*1) Simulation studies.* S2LAL circuits of substantial complexity should be designed and simulated using well-validated device models for a range of available processes. We hypothesize that, despite the increased complexity of S2LAL compared to other "fully" adiabatic logic styles such as 2LAL, it may ultimately prove capable of reaching much lower levels of energy dissipation per function in practice, due to the elimination of any sparking events that (in dynamic styles) might be caused upon reconnection of floating nodes which may have been subjected to, *e.g.*, capacitively induced voltage sag.

*2) Fabrication and power dissipation measurement of test chips*, to validate simulation results.

*3) Open-source hardware.* Due to the fully-static nature of S2LAL, it could be a relatively accessible technology for logic designers who are used to working in static CMOS, and thus are accustomed to not having to worry about whether floating nodes will sag or drift. So, S2LAL seems an appropriate target for producing an *open-source library* of reference cells and example designs for adiabatic CMOS, to encourage broader experimentation with adiabatic circuit concepts. To be accessible to the broadest community, such a library could ideally be developed for an open PDK and using open-source EDA tools. Environments of this nature are presently taking shape; one example being the open-source PDK release for SkyWater's 130 nm process; free shuttle runs based on this PDK and various open-source toolsets are currently being sponsored by Google and Efabless with outreach from FOSSI.[1]

*4) Cryogenic technologies.* In recent years, it has become apparent that modern CMOS technologies perform quite well at deep cryogenic temperatures (*e.g.*, ~4 K, and perhaps lower), spurring an increasing amount of research activity in cryogenic CMOS. In particular, at these temperatures, subthreshold slope becomes much steeper, and subthreshold leakage currents exponentially lower. This suggests the existence of a highly syn-ergistic opportunity to explore impactful applications for perfectly adiabatic logic styles such as S2LAL in the deep-cryogenic regime, for a couple of reasons:

*a) Ultra-low dissipation.* In-situ dissipation of a perfectly adiabatic logic style such as S2LAL is *only* limited by leakage. With subthreshold leakage becoming negligible at 4K, there is an opportunity to re-optimize the device structure to bring total leakage to such extreme low levels as to allow S2LAL to easily set a new record for energy efficiency of semiconductor-based logic; this can then potentially allow higher levels of *aggregate* parallel performance to be obtained in power-limited deep-cryo environments than has previously been possible.

*b) Power supply decoupling.* Highly efficient operation of adiabatic circuits requires precise power supply waveforms, and efficient operation at room temperature requires that these AC signals be supplied with high-$Q$ resonant energy recovery, which is quite challenging to achieve. But, for cryogenic applications, we can simply bring in supply lines from room-temperature sources, so the efficiency of the source is decoupled from the dissipation in the cryogenic environment. This can allow high-efficiency adiabatic circuits to be useful for *in situ* cryogenic applications, even before high-$Q$ resonant supplies become widely available.

*5) High-Q resonant supplies.* In a present project at Sandia, we are developing high-$Q$ resonant power supplies suitable for driving fully adiabatic circuits [19]; these can allow adiabatic logic styles such as S2LAL to be viable even for room-temperature applications. Completing this development effort and exercising our resonator designs in combination with S2LAL is another important direction to pursue. Also, in the longer term, design of appropriate *superconducting* resonators offers a way to eventually incorporate *extremely* high-$Q$ resonant supplies more directly within low-$T$ environments.

---

[1] See https://fossi-foundation.org/2020/06/30/skywater-pdk.

In conclusion, we have shown in detail how to construct a novel and highly perfected form of adiabatic CMOS that exhibits significant potential to enable practical applications and dramatic demonstrations which will illustrate the vast magnitude of future efficiency gains that can still be achieved for general digital computing.

## Acknowledgment

One of us (M. P. F.) would like to thank Mr. Richard Magnano, who helped inspire this innovation. On June 24th, when M. was creatively stalled in constructing S2LAL, R. showed M. the "Seed of Life" (right), an ancient sacred geometry figure, and indicated that others found it inspiring. Remotivated, M. resumed work on the problem, and determined within a day that this particular clocking scheme (Figs. 2, 6, and 7) with 8 out-of-phase cycles (echoing the 8 overlapping circles in the figure) in fact solves the problem and is as fast as possible.

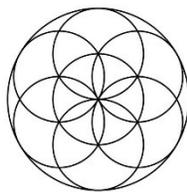

Thanks also are due to Erik DeBenedictis and Tom Conte for their encouragement in this work.